# Pressure-induced superconductivity in PrO$_x$FeAs

**Qingping Ding, Yangguang Shi, Hongbo Huang, Shaolong Tang, Shaoguang Yang***

Nanjing National Laboratory of Microstructures, Nanjing University, Nanjing 210093, China

*Corresponding author, e-mail: sgyang@nju.edu.cn

Superconductivity with $T_c$ of 45 K was realized in the tetragonal ZrCuSiAs-type PrO$_x$FeAs (x=0.75) under heat treatment of 1300 $^o$C and pressure of 6 GPa for 2 hours. Although the sample prepared at 900 $^o$C in vaccum possessed the same phase as that treated in pressure, the electrical transport measurement showed the similar behavior with its parent ReOFeAs (Re: rare-earth metal). This pressure-induced superconductivity could give us a hint to understand the origins of the newly-found iron based superconductors.

From the discovery of the superconductivity in F doped LaOFeAs [1], many efforts have been devoted to understanding the physics [2,3] and improving the properties of the iron based superconductors [4]. The superconducting transition temperature increases from 26 K to more than 50 K within short time [5-7]. Among these techniques, pressure-treated method is regarded as one of the most important techniques in the superconductor study. In fact, pressure has been used as an effective parameter to enhance or suppress the $T_c$ of superconductors in the iron-based superconductors [8-10]. Although insulator-metal phase transition in $NdO_{0.75}FeAs$ has been observed [11], pressure-induced superconductivity has not been reported. In this Letter, we report the superconductivity with $T_c$ = 45 K in $PrO_xFeAs$ (x=0.75) treated under the high pressure treatment.

The sample was prepared by the conventional solid state reaction method. PrAs, Fe and $Fe_2O_3$ were used as starting materials and weighed according to the chemical stoichiometry of $PrO_{0.75}FeAs$. In the sample preparation, the raw materials were grinded thoroughly and pressed into pellets. Then, the pellets were sealed in an evacuated quartz tube, and annealed at 900 $^o$C for 36 hours (sample A). The sample prepared in vacuum was further treated under a pressure of 6 GPa at the temperature of 1300 $^o$C for 2 hours (sample B). The following studies show that the high pressure treatment not only affects the crystal lattice parameters but also induces the sample to be a superconductor.

The phase structure of the two samples were characterized by powder X-ray diffraction (XRD) with Cu Kα radiation (λ=1.5418 Å) in the 2θ range of 20−80 degree with a step of 0.02 degree at room temperature. The XRD patterns of the two samples are shown in figure 1. For both samples, most peaks can be well indexed to tetragonal ZrCuSiAs-type structure with the space group P4/nmm. Small amount of $PrO_2$ impurities phase (maked with $^o$) can be observed in both samples, and some PrFeO phase (maked with *) exists in the sample after high pressure treatment. From the XRD results, it can be concluded that the tetragonal phase of $PrO_{0.75}FeAs$ is realized at the temperature of 900 $^o$C in vacuum. The peaks are magnified when finding out the detailed information from the XRD patterns. From the XRD results it can be found that the crystal structure has changed obviously after high pressure treatment. The insets of figure 1 are magnificationfrom the XRD patterns around peaks of (102) and (112), from which an obvious shift can be observed.

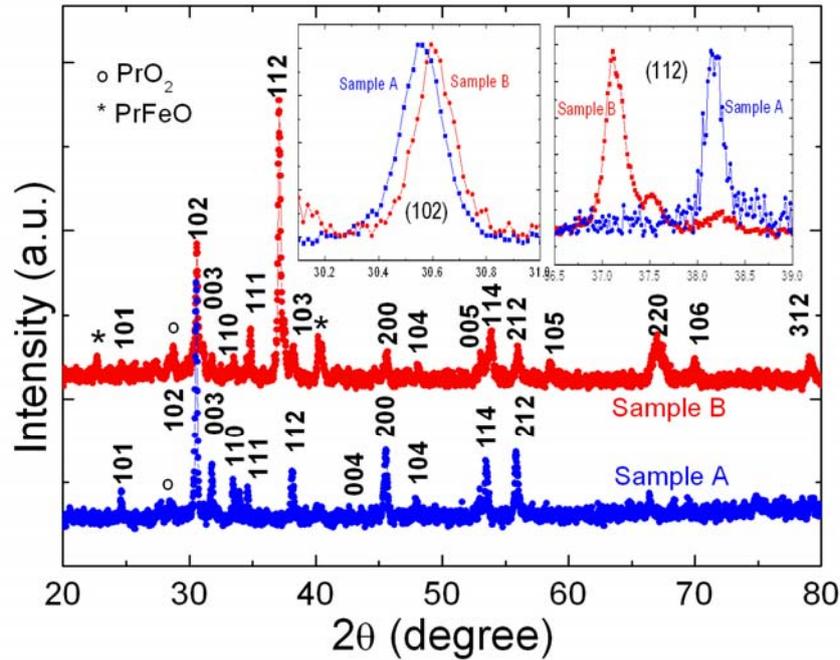

Fig.1 X-ray diffraction patterns of PrO$_{0.75}$FeAs before and after high pressure treatment. Insets are magnification of the two peaks (102) and (112) which shows us the pressure effect on the crystal lattice of the sample.

To investigate the electrical transport property, standard four-probe direct current (DC) resistivity measurements were performed on a physical property measurement system (PPMS). Figure 2 shows the DC resistivity dependence of the temperature under magnetic field of 0 and 1 T for the sample under the high pressure treatment. Inset is the result of the sample synthesized in vacuum, which is similar to the electrical transport property of its parent ReOFeAs. Compared with that of the vacuum-prepared sample, the resistivity decreased almost 2 orders at the normal state. This kind of decrease has been observed in the NdO$_{0.75}$FeAs system [11], which may be caused by pressure-induced increase in transfer interaction. Under zero field, the resistivity takes a fast drop from the temperature at 45 K, which refers to the onset $T_c$=45 K. The resistivity becomes immeasurable from 33 K with transition width of about 12 K. External magnetic field of 1 T makes the transition width broadening, but the onset transition temperature is not sensitive to this magnetic field, indicating that the upper critical field is very high for this superconductor.

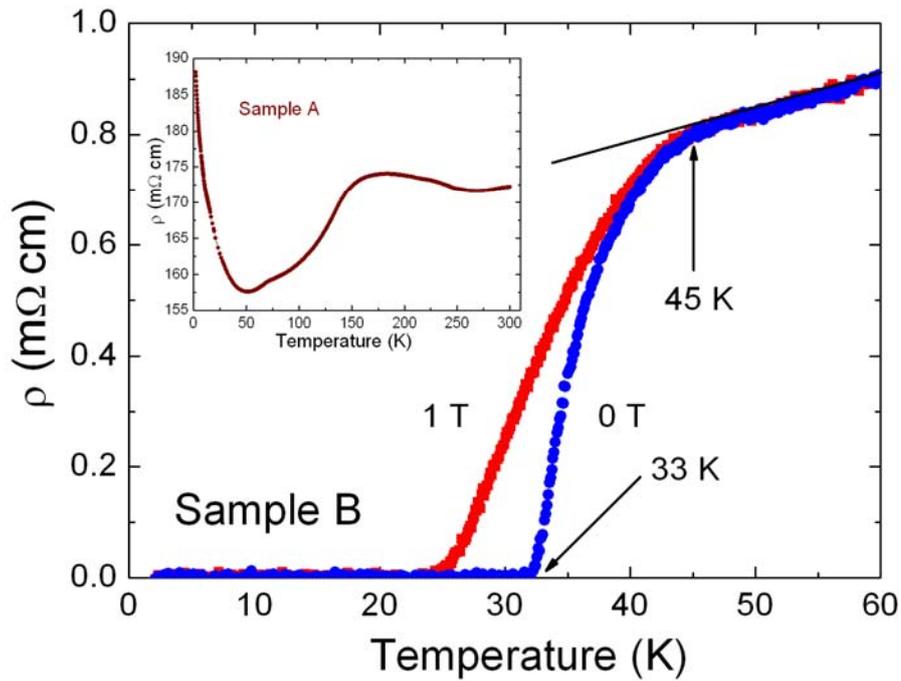

Fig.2 Resistivity dependence of the temperature of high pressure treated sample $NdO_{0.75}FeAs$ at 0 and 1 T. Inset is the result of the sample synthesized in vacuum.

The magnetization measurements of the high-pressure-treated sample were performed on a superconductor quantum interference device (SQUID). The alternate current (AC) susceptibility was measured under 1 Oe with the frequency of 10 Hz, and DC susceptibility was measured during warming cycle under 10 Oe after zero field cooling (ZFC) process. The temperature dependence of AC and DC susceptibility for the $PrO_{0.75}FeAs$ superconductor is shown in figure 3. The onset of the diamagnetic transition is at about 30 K. The magnetic measurements illustrate the Meissner effect in the high pressure-treated sample.

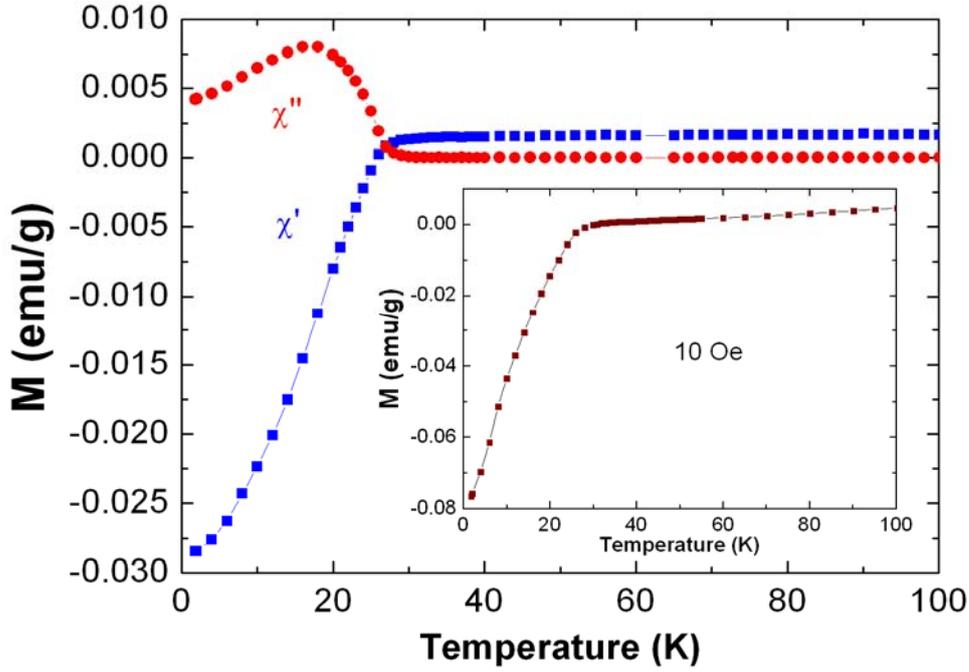

Fig.3 The temperature dependence of AC magnetic susceptibility of the high pressure treated superconducting $PrO_{0.75}FeAs$ sample. The inset is DC magnetic susceptibility after zero-field cooling.

It has been proved that the parent LaOFeAs takes both structural and magnetic phase transitions when the temperature decreases by neutron scattering, while F doping suppresses both of them, favoring the superconductivity [12]. The structural change is the possible reason for the pressure-induced superconductivity in $PrO_{0.75}FeAs$. In this work, the crystal lattice parameters of the samples before and after high pressure treatment are shown in table 1. The percentage of the constriction is calculated by using the data of the vacuum treated sample as the denominators, and minus means expansion. Accordingly, it can be concluded that most of the d-spacing values between the crystal planes are pressed to shrink except for (110) and (112) planes which expand after high pressure treatment. Among all of these d-spacing values, more attention should be paid to the variable of d-spacing value between (112) planes since it expands more than 2.8%, which is the largest one including constriction and expansion. The change in crystal lattice structure may be responsible for the superconductivity of the sample after the high pressure treatment.

Table 1 d-spacing parameters calculated from the XRD patterns (unit: nm)

| XRD peak | 102 | 003 | 110 | 111 | 112 | 200 | 114 | 212 |
|---|---|---|---|---|---|---|---|---|
| Sample A | 0.2926 | 0.2704 | 0.2569 | 0.2490 | 0.2357 | 0.1990 | 0.1714 | 0.1646 |
| Sample B | 0.2922 | 0.2703 | 0.2571 | 0.2473 | 0.2423 | 0.1986 | 0.1700 | 0.1641 |
| Constriction (%) | 0.1367 | 0.0037 | -0.0078 | 0.6827 | -2.800 | 0.2010 | 0.8154 | 0.3037 |

In summary, non-superconducting $PrO_{0.75}FeAs$ prepared at 900 $^oC$ in vacuum is changed to a superconductor with Tc of 45 K by the high pressure treatment at 6 GPa. XRD patterns show that the crystal structure has changed obviously after high pressure treatment. The resistivity takes a fast drop from the temperature at 45 K and it becomes immeasurable from 33 K with transition width of about 12 K. Meissner effect is observed below 30 k in the magnetic measurements.

**Acknowledgement:** This work was supported by the NSFC (10774068), NCET (07-0430) and "973" Program (2006CB921800).